\documentclass[aps,prl,reprint]{revtex4-1}
\usepackage{blindtext}
\usepackage{graphicx}
\usepackage{color}
\usepackage{amsmath}  
\usepackage{amssymb}
\usepackage{amsfonts} 
\usepackage{graphicx} 
\usepackage{xcolor}

\begin{document}
\title{Are models of local hidden variables for the singlet polarization state necessarily constrained by the Bell inequality ?}
\author{David H. Oaknin}
\email{d1306av@gmail.com}
\affiliation{Rafael Ltd, IL-31021 Haifa, Israel}

\begin{abstract}
The Bell inequality is thought to be a common constraint shared by all models of local hidden variables that aim to describe the entangled states of two qubits. Since the inequality is violated by the quantum mechanical description of these states, it purportedly allows distinguishing in an experimentally testable way the predictions of quantum mechanics from those of models of local hidden variables and, ultimately, ruling the latter out. In this paper, we show, however, that the models of local hidden variables constrained by the Bell inequality all share a subtle, though crucial, feature that is not required by fundamental physical principles and, hence, it might not be fulfilled in the actual experimental setup that tests the inequality. Indeed, the disputed feature neither can be properly implemented within the standard framework of quantum mechanics and it is even at odds with the fundamental principle of relativity. Namely, the proof of the inequality requires the existence of a preferred absolute frame of reference (supposedly provided by the lab) with respect to which the hidden properties of the entangled particles and the orientations of each one of the measurement devices that test them can be independently defined through a long sequence of realizations of the experiment. We notice, however, that while the relative orientation between the two measurement devices is a properly defined physical magnitude in every single realization of the experiment, their global rigid orientation with respect to a lab frame is a spurious gauge degree of freedom. Following this observation, we were able to explicitly build a model of local hidden variables that does not share the disputed feature and, hence, it is able to reproduce the predictions of quantum mechanics for the entangled states of two qubits.
\end{abstract}

\maketitle

{\bf 1.} The Bell theorem is one of the pillars upon which relies the widespread belief that quantum mechanics is the ultimate mathematical framework within which a hypothetical final theory of the fundamental building blocks of Nature and their interactions should be formulated. The theorem states through an experimentally testable inequality (the Bell inequality) that none theory of hidden variables that shares certain intuitive features can reproduce the predictions of quantum mechanics for the Bell states of two entangled qubits \cite{Bell}. In fact, since these predictions have been experimentally confirmed beyond any reasonable doubt \cite{Hansen,ScienceNews} all said generic models of hidden variables are currently ruled out.

In a Bell experiment a source emits pairs of particles whose polarizations are prepared in an entangled state:

\begin{equation}
\label{Bell_state}
| \Psi_{\Phi} \rangle = \frac{1}{\sqrt{2}} \left(|\uparrow \rangle^{(A)} \ |\downarrow\rangle^{(B)} - e^{-i \Phi} \ |\downarrow \rangle^{(A)} \ |\uparrow\rangle^{(B)}\right),
\end{equation}
where $\left\{|\uparrow\rangle, \ |\downarrow\rangle\right\}^{(A,B)}$ are single-particle eigenstates of Pauli operators $\sigma_Z^{(A,B)}$ along locally defined Z-axes, and two widely separated detectors oriented along independently set directions within the corresponding XY-planes test them. Upon detection each particle causes a binary response of its detector, either $+1$ or $-1$. Thus, each detected pair of entangled particles produces an outcome in the space of possible events $\left\{(-1,-1), (-1,+1), (+1,-1), (+1,+1)\right\}$. We refer to each detected pair as a single realization of the experiment. The experiment consists of a long sequence of realizations along which each one of the detectors can be switched between two possible settings, which we shall denote as $\Omega_A$ and $\Omega'_A$ for detector A and $\Omega_B$ and $\Omega'_B$ for detector B, defined with respect to local lab frames.
 
At the end of all these runs the outcomes recorded by the two detectors are compared and their statistical correlations computed at each one of the available settings. Quantum mechanics predicts, and experimental tests confirm, that these correlations are given by

\begin{equation}
\label{correlation}
E(\Delta-\Phi) = -\cos\left(\Delta - \Phi\right),
\end{equation}
where $\Delta$ is the relative angle between the orientations of the two measuring devices and the phase $\Phi$ is defined by (\ref{Bell_state}). Therefore, it can be readily check that
\begin{equation}
\label{the_fact}
\left|E(+\pi/4) + E(-\pi/4) + E(-\pi/4) - E(-3\pi/4) \right| = 2\sqrt{2}.
\end{equation}
On the other hand, the CHSH version of the Bell inequality states that for all models of hidden variables that share certain intuitive features the following inequality
\begin{eqnarray}
\label{CHSH_original}
\left|E(\Delta_1) + E(\Delta_2) + E(\Delta_1-\delta) - E(\Delta_2-\delta)\right| \le 2,
\end{eqnarray}   
must hold for any set of values $\left(\Delta_1,\Delta_2,\delta\right)$ and, in particular, for $\Delta_1=+\pi/4$, $\Delta_2=-\pi/4$ and $\delta=+\pi/2$ \cite{CHSH}. Since experiments have confirmed the predictions of quantum mechanics (\ref{the_fact}) beyond any reasonable doubt, all models of hidden variables constrained by the inequality (\ref{CHSH_original}) are experimentally ruled out. 

This indisputable conclusion is widely interpreted as an experimentally verified proof of the impossibility to describe quantum phenomena within the framework of any underlying model of local hidden variables. Nonetheless, the right statement is that it is impossible to describe quantum phenomena within the framework of any model of hidden variables that shares the intuitive features assumed by the Bell theorem. In fact, as we show below all the models of hidden variables constrained by the CHSH inequality (\ref{CHSH_original}) share a subtle, though crucial, feature that is not required by fundamental physical principles and, hence, it might not be fulfilled in the actual experimental setup that tests the inequality. Indeed, we have shown in \cite{david,david2} that once the disputed assumption is lifted it is straightfoward to build an explicitly local model of hidden variables that reproduces the predictions of quantum mechanics for the Bell states.

Let us start our discussion about the Bell experiment with the following observation: within the standard framework of quantum mechanics the source of entangled particles cannot be properly described with respect to a lab frame, but only with respect to a reference setting of the two detectors. The argument goes as follows. The Bell states (\ref{Bell_state}) are defined in terms of the bases $\{|\uparrow \rangle, \ |\downarrow \rangle\}^{(A,B)}$ of single-particle eigenstates of the Pauli operators $\sigma_Z^{(A,B)}$. Since these eiegenstates are defined up to a global phase, the phase $\Phi$ in (\ref{Bell_state}) could not be properly defined with respect to a lab frame of reference. In order to properly define this phase and, hence, the source of entangled particles we must choose an arbitrary reference setting of the two measurement devices. The phase $\Phi$ is then defined with respect to this reference setting of the detectors with the help of the measured correlations between their outcomes, $E=-\cos(\Phi)$. We can then use this reference setting to properly define also a relative rotation $\Delta$ of the orientations of the two apparatus. It is interesting to notice at this point that the definitions of the phase $\Phi$ and the angle $\Delta$ do not rely at all on the quantum formalism and, therefore, we shall use the same definitions later on to build our model of hidden variables to describe the experiment. It is also important to notice that since we must use an otherwise arbitrary setting of the detectors as a reference in order to properly describe the experiment we cannot in any proper sense define their global rigid orientation: it is an spurious gauge degree of freedom. 
  
Nevertheless, the proof of the CHSH inequality does not properly recognize this spurious gauge degree of freedom. It proceeds as follows. Let us label as $\left\{\lambda\right\}_{\lambda \in {\cal S}}$ the space of all possible hidden configurations of the pair of entangled particles and let $\rho(\lambda)$ be the (density of) probability of each one of them to occur in every single realization of the experiment. It is then assumed that it is possible to assign to each possible configuration $\lambda \in {\cal S}$ a 4-tuple of binary values 

\begin{equation}
\label{Bassumption}
\left(s^{(A)}_{\Omega_A}(\lambda), s^{(A)}_{\Omega'_A}(\lambda), s^{(B)}_{\Omega_B}(\lambda), s^{(B)}_{\Omega'_B}(\lambda)\right) \in \left\{-1,+1\right\}^4
\end{equation}
to describe the outcomes that would be obtained at each one of the measurement devices in case that their orientations were set along each one of the two available settings - $\Omega_A$, $\Omega'_A$ and $\Omega_B$, $\Omega'_B$ - defined with respect to local lab frames. Under this assumption, which we shall refer to as the Bell assumption, it is straightforward to show that for all possible configurations $\lambda \in {\cal S}$, 

\begin{eqnarray}
\label{CHSH_proof}
\nonumber
s^{(A)}_{\Omega_A}(\lambda)\cdot \left(s^{(B)}_{\Omega_B}(\lambda) + s^{(B)}_{\Omega'_B}(\lambda)\right)+\hspace{0.7in}\\ 
+ s^{(A)}_{\Omega'_A}(\lambda)\cdot \left(s^{(B)}_{\Omega_B}(\lambda) - s^{(B)}_{\Omega'_B}(\lambda)\right)=\pm2,
\end{eqnarray}
since the first term equals either $+2$ or $-2$ when $s^{(B)}_{\Omega_B}(\lambda)$ and $s^{(B)}_{\Omega'_B}(\lambda)$ have the same sign and equals $0$ when they have opposite signs, while the second term equals $0$ when $s^{(B)}_{\Omega_B}(\lambda)$ and $s^{(B)}_{\Omega'_B}(\lambda)$ have the same sign and equals either $+2$ or $-2$ when they have opposite signs. The CHSH inequality (\ref{CHSH_original}) is then obtained by averaging this magnitude over the whole universe of events ${\cal S}$ and noticing that each one of the four terms in the integrand produces one of the required correlations:

\begin{eqnarray}
\label{CHSH_integral}
\nonumber
-2 \le \int d\lambda \ \rho(\lambda) \cdot \ \left[s^{(A)}_{\Omega_A}(\lambda)\cdot \left(s^{(B)}_{\Omega_B}(\lambda) + s^{(B)}_{\Omega'_B}(\lambda)\right)\right.+\hspace{0.3in}\\ 
\nonumber
+ \left. s^{(A)}_{\Omega'_A}(\lambda)\cdot \left(s^{(B)}_{\Omega_B}(\lambda) - s^{(B)}_{\Omega'_B}(\lambda)\right)\right] \le +2.
\end{eqnarray}

The Bell assumption (\ref{Bassumption}) intuitively seems a trivial feature of any model of local hidden variables, which seemingly simply states that the response of each detector to each possible hidden configuration $\lambda \in {\cal S}$ does not depend on the orientation chosen for the other detector. Indeed, this assumption would be indisputable if each particle of every single entangled pair could be tested at once along the two available orientations of its detector. However, since each particle of every entangled pair can be actually tested along only one possible orientation of its detector it is crucial to identify the actual physical degrees of freedom of the experimental set-up. In fact, as we shall now show the assumption (\ref{Bassumption}) is not required by fundamental physical principles and, therefore, might not be fulfilled in the actual experiments that test the Bell's inequality.

In general, we should allow for each one of the two detectors to define its proper set of coordinates over the space ${\cal S}$ of possible hidden configurations. Thus, let us denote as $\lambda_A$ and $\lambda_B$ the two sets of coordinates associated to detectors A and B, respectively. Since these two sets of coordinates parameterize the same space of hidden configurations ${\cal S}$ there must exist some invertible transformation that relates them:
\begin{eqnarray}
\label{Oaknin}
\lambda_B & = & -{\cal L}(\lambda_A; \ \Delta - \Phi), 
\end{eqnarray}
which may depend parametrically on the relative angle $\Delta - \Phi$ between the orientations of the two detectors. This transformation must fulfill the constraint
\begin{equation}
\label{freewill}
d\lambda_A \ \rho(\lambda_A) = d\lambda_B \ \rho(\lambda_B),
\end{equation}
which simply states that the probability to occur of every hidden configuration must remain invariant under a change of coordinates, while the density of probability $\rho(l), \ l \in [-\pi, \pi)$, is functionally invariant for the two sets of coordinates. It can be readily shown \cite{david,david2} that 
when we define 
\begin{equation}
\label{dens}
\rho(l) = \frac{1}{4}\left|\sin(l)\right|
\end{equation}
the constraint (\ref{freewill}) directly leads to the correlation (\ref{correlation}).

Furthermore, the transformation law (\ref{Oaknin}) complies with the trivial demand that a relative rotation of the measuring devices by an angle $\Delta$ followed by a second relative rotation by an angle $\Delta'$ results into a final rotation by an angle $\Delta+ \Delta'$ with respect to the original reference setting. This can be readily shown as follows. Consider, for example, a setting in which the angular coordinates of the hidden configurations with respect to each one of the two measurement devices, $\lambda_A$ and $\lambda_B$, are related by the transformation 
\begin{eqnarray}
\label{T1}
\lambda_B & = & -{\cal L}(\lambda_A; \ \Delta), 
\end{eqnarray}
Thus, with respect to this setting the source is described by a phase $\Phi=-\Delta$. Hence, by adding a relative angle $\Delta'$ to the relative orientation of the detectors we obtain a new setting in which the sets of coordinates associated to the two detectors are related by the transformation 
\begin{eqnarray}
\label{L0}
\lambda'_B =-L(\lambda_A; \Delta'-\Phi)=-L(\lambda_A; \Delta'+\Delta).
\end{eqnarray} 
By comparing with the transformation law (\ref{T1}) we realize that the final setting corresponds to a relative angle of $\Delta + \Delta'$ with respect to the original reference setting, as we had demanded.

Finally, since the global rigid orientation of the two devices is an spurious gauge degree of freedom, the set of coordinates over the space of hidden configurations may accumulate a non-zero geometric phase through a cyclic transformation:
\begin{equation}
\label{Gphase}
\left(-{\cal L}_{\Delta_2}\right) \circ \left(-{\cal L}_{\Delta_2 - \delta}\right) \circ \left(-{\cal L}_{\Delta_1 - \delta}\right) \circ \left(-{\cal L}_{\Delta_1}\right) = {\cal L}_{\alpha} \neq \mathbb{I}.
\end{equation}
The appearance of geometric phases in physical models involving gauge symmetries is a well-known phenomenon \cite{Wilczek} and, therefore, we should not rule out the possibility that it also occurs in models of hidden variables that describe quantum phenomena. Under such circumstances there does not exist a common set of coordinates in which we can jointly define binary responses for the two detectors in each one of their two available orientations. Hence, the Bell's assumption (\ref{Bassumption}) does not hold and, therefore, such models are not constrained by the inequality (\ref{CHSH_proof}). 

In the presence of a non-zero geometric phase we must choose the orientation of one of the detectors as a common reference direction in order to compare the four experiments involved in the CHSH inequality. Thus, instead of (\ref{CHSH_proof}) we should have written:
\begin{eqnarray}
\label{CHSH_new}
s(\lambda_A) \cdot \left[s(\lambda_B) + s(\lambda'_B) + s(\lambda''_B) - s(\lambda'''_B)\right],
\end{eqnarray}
where we have assumed that the two detectors share the same universal binary response function $s(l), l \in [-\pi,\pi)$, and we have now defined
\begin{eqnarray}
\label{T2}
\lambda_B & = & -{\cal L}(\lambda_A; \ \Delta_1), \\
\lambda'_B & = & -{\cal L}(\lambda_A; \ \Delta_2), \\
\lambda''_B & = & -{\cal L}(\lambda_A; \ \Delta_1 - \delta), \\
\lambda'''_B & = & -{\cal L}(\lambda_A; \ \Delta_2 - \delta).
\end{eqnarray}
It is obvious that the new expression (\ref{CHSH_new}) is no longer constrained to equal either $+2$ or $-2$. 

This requirement can be understood as follows. Once we accept that the hidden polarization properties of the entangled particles need to be defined with respect to the reference frame set by the orientation of the detectors that test them, and that the descriptions associated with different orientations of the detectors are related by gauge transformations, we can allow that in the description referred to orientation $\Omega_A$ of detector A the polarization properties of its particle along any other direction $\Omega'_A$ would not be binary. Of course, in the description referred to orientation $\Omega'_A$ the polarization properties along this direction must be binary, while the properties along $\Omega_A$ would not be necessarily so. In other words, the hidden polarization properties of the entangled particles may not be scalar magnitudes under a rotation of the detector that test them, much like the components of an electromagnetic field are not scalar magnitudes under the transformation that connects two observers related by a boost.   

Finally, it is worth to stress that eq. (\ref{Oaknin}), which relates the sets of coordinates defined by each one of the two detectors in a Bell experiment, does not introduce any non-local interaction between them. In order to clarify this issue let us consider a source that produces pairs of parallel macroscopic arrows randomly oriented along a locally defined $XY$ plane. The arrows are then parallely transported in opposite directions along the $Z$ axis to two distant detectors, each one of them consisting of an arrow that can also be arbitrarily oriented in the $XY$ plane. For every pair of arrows the following constraint is fulfilled:
\begin{equation} 
\label{euclid}
\theta_A = \theta_B - \Theta,        
\end{equation}
where $\theta_A$ is the relative angle between the orientation of detector A and its incoming arrow, $\theta_B$ is the relative angle between the orientation of detector B and its incoming arrow and $\Theta$ is the relative angle between the orientations of the two detectors. The constraint (\ref{euclid}) is dictated by the euclidean structure of the macroscopic space. Thus, it is fulfilled no matter who decides how to orient the detectors or whenever the decisions are taken and, obviously, it does not introduce any non-local interaction between the detectors. Eq. (\ref{Oaknin}) is nothing but a non-linear generalization of the euclidean relationship (\ref{euclid}), and it simply means that the entangled particles may carry with them a non-euclidean metric. 

In summary, we have shown that the Bell theorem holds only for a very particular class of models of local hidden variables that share a subtle, though crucial, feature. This feature, nonetheless, is not required by fundamental physical principles and it is not necessarily fulfilled in the actual experimental setup that tests the inequality. Indeed, following this observation we have presented in \cite{david,david2} an explicitly local statistical model of hidden variables that does not share the said feature and reproduces the predictions of quantum mechanics for the Bell states.


\begin{references}


\bibitem{Bell} J.S.~Bell, Physics {\bf 1}, 195-200 (1964).

\bibitem{Hansen} B.~Hensen {\it et al}, Nature {\bf 526}, 682 (2015).

\bibitem{ScienceNews} H.~Wiseman, Nature {\bf 526}, 649 (2015).


\bibitem{CHSH} J.F.~Clauser, M.A.~Horne, A.~Shimony and R.A.~Holt, Phys. Rev. Lett. {\bf 23}, 880–884 (1969).

\bibitem{david} D.H.~Oaknin, Frontiers in Physics 8:142 (2020).

\bibitem{david2} D.H.~Oaknin, arXiv:1411.5704.

\bibitem{Wilczek}  F.~Wilczek and A.~Shapere, eds. (1989). Geometric Phases in Physics. Singapore: World Scientific.
































\end{references}
\end{document}